\documentclass[showpacs,amssymb,superscriptaddress,notitlepage,nofootinbib,preprint,longbibliography]{revtex4-2}
\usepackage{graphicx}
\usepackage{amsmath}
\usepackage{amsfonts}
\usepackage{bm}
\usepackage[colorlinks=true]{hyperref}
\usepackage[all]{hypcap}
\usepackage[utf8]{inputenc}
\usepackage{url}

\usepackage{float}
\usepackage{xcolor}
\usepackage{multirow}
\usepackage{mathtools}
\usepackage{tikz}

\hypersetup{
    colorlinks,
    citecolor=[rgb]{0.16,0.53,0.80},
    linkcolor=[rgb]{0.16,0.03,0.50},
}

\usepackage{graphicx,color}

\def\be{\begin{equation}}
\def\ee{\end{equation}}
\def\bea{\begin{eqnarray}}
\def\eea{\end{eqnarray}}
\def\ptl{\partial}

\begin{document}

\thispagestyle{empty}

\centerline{\Large{ \bf Scalar Product for a Version of
Minisuperspace Model }} \centerline{\Large{\bf  with Grassmann
Variables }}
\medskip
\centerline{S.L. Cherkas $^a$, V.L. Kalashnikov$^b$}
\medskip
\centerline{$^a$\it Institute for Nuclear Problems, Belarus State
University} \centerline{\it Minsk 220006, Belarus}
\centerline{cherkas@inp.bsu.by}
\medskip
\centerline{$^b$\it  Department of Physics,}\centerline{\it
Norwegian University of Science and Technology,}\centerline{\it
H{\o}gskoleringen 5, Realfagbygget, 7491, Trondheim, Norway}
\centerline{vladimir.kalashnikov@ntnu.no}
\medskip
\centerline{\small Submitted: December 8, 2023}

\smallskip

{\fontsize{10}{11}\selectfont Grassmann variables are used to
formally transform a system with constraints into an unconstrained
system. As a result, the Schr\"{o}dinger equation arises instead
of the Wheeler--DeWitt one. The Schr\"{o}dinger equation describes
a system's evolution, but a definition of the scalar product is
needed to calculate the mean values of the operators. We suggest
an explicit formula for the scalar product related to the
Klein--Gordon scalar product. The calculation of the mean values
is compared with an etalon method in which a redundant degree of
freedom is excluded. Nevertheless, we note that a complete
correspondence with the etalon picture is not found. Apparently,
the picture with Grassmann variables requires a further
understanding of the underlying Hilbert space.}

\smallskip

{\fontsize{10}{11}\selectfont {\bf Keywords:} minisuperspace
model; quantum evolution; ghost variables; operator mean values}

\smallskip

\section{Introduction.}

There is a principal possibility to construct the theory of
quantum gravity (QG) from the point of view that gravity is a
usual physical system with constraints~\cite{Gitman,Henneaux}. As~
a result, it has to be quantized using Dirac
brackets~\cite{Dirac}.

The physical question arises: what type of gravity theory must be
quantized? It is hardly general relativity (GR), because GR
suffers from the loss of information (unitarity) in black
holes~\cite{PhysRevHawk,Giddings2006} (however, see,
e.g.,~\cite{Page1993,penington2020entanglement,Hashimoto2020}) and
from the vacuum energy
problem~\cite{akhmedov2002vacuum,particles1010010,barv18}. It
seems possible~\cite{ch1} to repair GR by restricting it to a
class of submanifolds without black
holes~\cite{cherkas2020eicheons,carb2023}. Simultaneously,
the~possibility of arbitrarily choosing an energy density level
 appears~\cite{ch1,Haridasu}, which removes the vacuum energy problem,
at least for massless particles. Contributions of the masses of
particles into vacuum energy density $\sim M_p^2m^2$ have to be
mutually compensated for~\cite{Visser2019}. Contributions of the
order of $\sim m^4$ also have to be compensated for but~by taking
condensates into account~\cite{Haridasu}.   The~resulting theory,
which has  features such as partial gauge fixing, preferred
reference frame, and~the existence of \ae
ther~\cite{Townsend,Cherkas2022} (in the form  of condensates and
vacuum polarization) could be a suitable candidate
for~quantization.

Another (mathematical) question is how to realize the commutation
relations corresponding to Dirac brackets, which again implies
gauge fixing by defining auxiliary conditions to convert a system
from the first class to the second class. So far, there is no
constructive way to do so generally~\cite{concept}. In~particular,
the~implementation of Dirac brackets in 3 + 1 GR has not yet been
achieved. Even for  2 + 1 gravity, Dirac brackets have a rather
complicated structure~\cite{Meusburger2011}. Moreover, in~the case
of gravity quantization, the~gauge-fixing conditions must be
time-dependent to introduce time into the theory. Instead, one
could use the quasi-Heisenberg
picture~\cite{Cher2005,Cher2012,Cher2013,Cher2017}, where the
commutator relations corresponding to Dirac brackets are
determined at some fixed moment of time, e.g.,~near a small-scale
factor that simplifies a~problem.

A more radical method is introducing Grassmann
variables~\cite{Faddeev1974,Faddeev,Savchenko2004,Vereshkov2013,Upadhyay2015,Chauhan2022},
which formally
 reduces a system with constraints
to an unconstrained one.  However, if~one applies Grassmann
variables to calculate not only the scattering amplitudes but also
the mean values of operators, the~questions about Hilbert spaces
and scalar products arise~\cite{ruffini2005,Kleefeld2006,Mont}.

For simplicity, the~question about scalar products could be
considered in a minisuperspace model example. Minisuperspace
models  represent examples of simple systems with  constraints and
are widely
used~\cite{Rubak,Lemos1996,PhysRevD123512,Gryb2019,Gielen2020,Gielen2022,mena,Bojowald2015,balcerzak2023spinor,Kaya2023}
to understand the main features of gravity quantization.
Without~experimental data for the minisuperspace model, one would
not be able to straightforwardly check
 different approaches to gravity quantization. Fortunately, an~etalon quantization method for the minisuperspace model that  ``could not be
 wrong'' exists. It consists of the explicit exclusion  of the redundant
degree of freedom, initially, to obtain a physical
Hamiltonian~\cite{Barv2014,cherkas2020evidence} through explicit
gauge fixing. Considering this ``etalon'' method  implies that QG
has to
violate gauge invariance
. Certainly, opposite points of view exist; e.g.,~loop quantum
gravity promises ``covariant with respect to diffeomorphisms''
QG~\cite{Ashtekar2021} (however, see some comments
in~\cite{Bojowald2021}).

\section{The Etalon Picture with the Exclusion of the Redundant Degrees of~Freedom}
\label{etal}

Let us consider the action functional of gravity  minimally
coupled to a massless scalar field:
\bea
S=\frac{1}{16\pi G}\int R  \sqrt{-g}\,d^4x
+\frac{1}{2}\int\ptl_\mu\phi\,
g^{\mu\nu}\ptl_\nu\phi\sqrt{-g}\,d^4x,
\label{lag}
\eea
where $R$ is  the scalar curvature; Greek indexes run from 0 to 3;
$G$ is the Newton constant; $g_{\mu \nu}$ is the metric tensor,
with g being its determinant. Considering a uniform, isotropic,
and~flat universe
\be
ds^2=g_{\mu\nu}dx^\mu dx^\nu=a^2(N^2d\eta^2-d^2\bm r),
\ee
where functions $a$ and $N$ depend only on $\eta$, reduces action
(\ref{lag}) to
\be
S=\frac{1}{2}\int\frac{1}{N}\left(-{M_p^2}a^{\prime
2}+a^2\phi^{\prime 2}\right)d \eta=\int \left(-p_a
a^\prime+\pi_\phi\phi^\prime- N\left(-\frac{1}{2
}p_a^2+\frac{\pi_\phi^2}{2a^2}\right)\right)d \eta,
\label{lag1}
\ee
where the reduced Planck mass, $M_p=\sqrt{\frac{3}{4\pi G}}$, is
used, which is set to unity everywhere  further  for simplicity.
The~Hamiltonian
\be
H=N\left(-\frac{1}{2 }p_a^2+\frac{\pi_\phi^2}{2a^2}\right),
\label{phihamil}
\ee
also determines the Hamiltonian constraint
\be
 \Phi_1=-\frac{1}{2
}p_a^2+\frac{\pi_\phi^2}{2a^2}=0,
\ee
by virtue of $\frac{\delta S}{\delta N}=0$. The~time evolution of
an arbitrary observable $A$ is expressed with the Poisson brackets
\be
\frac{dA}{d\eta}= \frac{\ptl A}{\ptl \eta}+\{H,A\},
\ee
which  are defined~as

\be
 \{A,B\}=\frac{\ptl A}{\ptl \pi_\phi}\frac{\ptl
B}{\ptl \phi}-\frac{\ptl A}{\ptl \phi}\frac{\ptl B}{\ptl
\pi_\phi}-\frac{\ptl A}{\ptl p_a}\frac{\ptl B}{\ptl a}+\frac{\ptl
A}{\ptl a}\frac{\ptl B}{\ptl p_a}.
\ee

\noindent  {The full system of equations} 
 of motion has the form
 \bea
  \phi^\prime=\frac{\ptl H}{\ptl \pi_\phi}=N\,\frac{\pi_\phi}{a^2},~~~\pi_\phi^\prime=-\frac{\ptl H}{\ptl\phi }=0,~~\Longrightarrow~~\pi_\phi=k=const,\label{eqforphi}\\
a^\prime=-\frac{\ptl H}{\ptl p_a}=N\,p_a,~~~
 p_a^\prime=\frac{\ptl H}{\ptl a}=-\frac{N\,k^2}{a^3}.
 \label{theeqm}
 \eea

\noindent  {The additional time-dependent }gauge-fixing condition
\be
\Phi_2=a-\sqrt{2|\pi_\phi| \eta}
\label{f2}
\ee can be introduced
as the constraint $\Phi_2$, which fixes  $N$ to be equal unity.
The condition (\ref{f2}) also fixes the direction of time because
the scale factor increases with time. The~solutions of
Equation~(\ref{theeqm}) are
\be
a=\sqrt{2|\pi_\phi| \eta},
~~~~~~~~~~~~\phi=\frac{\pi_\phi}{2|\pi_\phi|}\ln\eta +const.
\ee
 {The constraints }$\Phi_1$, $\Phi_2$ allow reducing this simple
system to a sole degree of freedom. Let us make some general notes
about the exclusion of the variables and coming to the physical
Hamiltonian. Let  system variables be $Q_1,\ldots,
Q_m,q_{m+1},\ldots, q_n$, $P_1,\ldots, P_m,p_{m+1},\ldots, p_n$
and one would like to exclude $n-m$ coordinates $q_k$ and
momentums $p_k$ using the constraints and additional conditions
$\Phi_k$, to~have only the variables $Q_k$ and momentums $P_k$
finally. Equating the
Lagrangians before and after  {exclusion leads to} 
\be
\sum_{k=1}^{m}P_k\,dQ_k+\sum_{k=m+1}^{n}p_k\,dq_k-\sum_{k=m+1}^{m}\lambda_k
\Phi_k=\sum_{k=1}^{m}P_k\,dQ_k-H_{phys}(\bm Q,\bm P)d\tau+d F(\bm
Q,\bm P,\tau),
\label{constrexc}
\ee
where the left-hand part of (\ref{constrexc}) contains all the
constraints, including the Hamiltonian one, whereas the right-hand
side contains the full differential of some function $F(\bm Q,\bm
P,\tau)$. At the constraint surface $\Phi_k$,
Equation~(\ref{constrexc}) reduces to
\be
H_{phys}=-\sum_{k=m+1}^{n}p_k\,q_k^\prime+\frac{d F(\bm Q,\bm
P,\tau)}{d\tau}.
\label{hpwf}
\ee
 {The function} $F$ has to be chosen in such a way that
$H_{phys}$ reproduces correct equations of motion by
\be
\bm Q^\prime=\frac{\ptl H_{phys}}{\ptl \bm P}, ~~~~~\bm
P^\prime=-\frac{\ptl H_{phys}}{\ptl \bm Q}.
\label{testeqmov}
\ee

 Let us take $\pi_\phi$ and $\phi$ as
 physical variables, then $a$ and $p_a$ have to be excluded by
the constraints. Substituting $p_a$, $a^\prime$ and $a$ into
(\ref{lag1}) results in
\be
L=\int \left(\pi_\phi\phi^\prime
-H_{phys}(\phi,\pi_\phi,\eta)\right)d\eta,
\label{L2}
\ee
where
\be
H_{phys}(\phi,\pi_\phi,\eta)=p_a
a^\prime=\frac{|\pi_\phi|}{2\,\eta}.
\label{hph}
\ee
 {Hamiltonian }(\ref{hph}) reproduces the equations of motion
(\ref{eqforphi}) correctly; thus, function $F$ in (\ref{hpwf})
equals zero for this  {particular case}
\footnote{For instance, in~ a more general case,
$H={N}\left(-\frac{p_a^2}{2}+\frac{\pi_\phi^2}{2a^n}\right)$, and~
the conserved in time gauge-fixing condition
$a-\left((1+\frac{n}{2})k\,\eta\right)^{\frac{2}{n+2}}$=0,
physical Hamiltonian
$H_{phys}=(n+2)^{\frac{2-n}{2+n}}\left(\frac{k^4}{16\eta^{2n}}\right)^{\frac{1}{n+2}}$
and $\frac{dF}{d\eta}=\frac{n-2}{n+2}H_{phys}$.}.

The most simple and straightforward way to describe quantum
evolution is to formulate the Schr\"{o}dinger equation
\be
i\ptl_\eta \Psi=\hat H_{phys}\Psi
\label{sh}
\ee
with the physical Hamiltonian (\ref{hph}). In~the momentum
representation, the~operators become
\be
\hat \pi_\phi=k, ~~~~~\hat \phi=i\frac{\ptl}{\ptl k}.
\label{repr}
\ee

\noindent  {The solution of} Equation~(\ref{sh}) is written as
\be
\Psi(k,\eta)=C(k)\left(\frac{2}{e}|k|\eta\right)^{-i|k|/2}.
\label{psi}
\ee

It is possible to calculate the mean values of an arbitrary
operator $\hat A(k,i\frac{\ptl}{\ptl k})$ built from $\hat
\phi=i\frac{\ptl}{\ptl k}$ and $a=\sqrt{2|k|\eta}$  with respect
to a  wave packet $C(k)$ in the following way
\be
<C|\hat A|C>=\int \Psi^*(k,\eta)\hat A\,\Psi(k,\eta)dk.
\label{hata}
\ee

 Since the basic wave function $\left(\frac{2}{e}|k|\eta\right)^{-i|k|/2}$ contains a
  module of $k$, a~singularity may arise at $k=0$ if $\hat A$
 contains degrees of
the differential operator $\frac{\ptl}{\ptl k}$. That may violate
hermicity. To~avoid this, the~wave packet has to be turned to zero
at $k=0$. For~instance, it could be taken in the Gaussian form
\be
C(k)=\frac{2 \sigma^{-5/2}}{\sqrt 3 \pi^{1/4}}k^2\exp(-
k^2/(2\sigma^2))
\label{pak}
\ee
with the multiplier $k^2$ in the front of the~exponent.

 Let us come to the calculation of some mean values
taking the parameter $\sigma=1$. The~mean value of $a^2$
is\footnote{The mean value of $a^2$ is singular at $\eta=0$.
Moreover, one may consider that the singularity stores information
about the quantum state defined by the wave packet $C(k)$
(see~\cite{Cherkas2022b} for a general discussion). On~the other
hand, there is a ``no-boundary'' proposal for  a non-singular
origin of the universe (for a review, see~\cite{Lehners2023}).  }
\be
<C|a^2|C>=\frac{16}{3\sqrt{\pi}} {\eta}\int_{0}^\infty e^{-k^2}
k^5 dk=\frac{16}{3\sqrt{\pi}}
\eta.~~~~~~~~~~~~~~~~~~~\label{resa2}\\
\ee
 {The next quantity }is expressed as
\be
<C|a^4|C>=\frac{16}{3\sqrt{\pi}} {\eta}^2\int_{-\infty}^\infty
e^{-k^2} k^6 dk=10\eta^2.~~~~~~~~~~~~~~~~~~~\label{resa4}\\
\ee

\noindent  {Other mean values} for this wave packet were
calculated in~\cite{cherkas2020evidence}.

\section{ Evolution in Extended~Space}
Another derivation of the physical Hamiltonian (\ref{hph}) is
given by  the continual integrals  considering the transition
amplitude from \emph{ {in} 
} to \emph{ {out}} states. Using canonical gauge fixing condition
$\Phi_2=a-\sqrt{2\eta|\pi_\phi|}$ leads to\vspace{-6pt}

\bea
<out|in>=Z=\int e^{i\int\left(\pi_{\phi}\phi^\prime -p_a
a^\prime-N\left(-\frac{1}{2}p_a^2+\frac{\pi_{\phi}^2}{2a^2}\right)\right)d\eta}\Pi_\eta
\Delta_{FP}\Pi_\eta \delta\bigl(\Phi_2\bigr)\mathcal D N\mathcal D
p_a \mathcal D a\mathcal D \pi_\phi \mathcal D \phi=\nonumber\\
\int e^{i\int\left(\pi_{\phi}\phi^\prime -p_a
a^\prime\right)d\eta}\Pi_\eta
p_a\Pi_\eta\delta\left(-\frac{1}{2}p_a^2+\frac{\pi_{\phi}^2}{2a^2}\right)\Pi_\eta
\delta\bigl(a-\sqrt{2\eta|\pi_\phi|}\bigr)\mathcal D p_a \mathcal
D a\mathcal D \pi_\phi \mathcal D
\phi=~~~\nonumber\\
\int  e^{i\int\left(\pi_{\phi}\phi^\prime
-|\pi_\phi|/(2\eta)\right)d\eta} D \pi_\phi \mathcal D\phi.~~~
\label{deriveh}
\eea

 Under the derivation of (\ref{deriveh}), it was used that
$\Delta_{FP}=\{\Phi_1,\Phi_2\}=p_a$, and~$\Pi_\eta
\delta\left(-\frac{1}{2}p_a^2+\frac{\pi_{\phi}^2}{2a^2}\right)=\frac{\Pi_\eta\delta\left(p_a-|\pi_\phi|/a
\right)}{\Pi_\eta p_a}$. From~(\ref{deriveh}) follows the formula
(\ref{hph}) for a physical Hamiltonian.   A~principle of
derivation is clear: obtaining an expression of a kind $\int
e^{i\int\left(\sum_i p_i q^\prime_i -H(\bm p,\bm q)\right)d\eta}
\mathcal D \bm p \mathcal D \bm q$ without any  pre-exponential
factors and extracting $H(\bm p,\bm q)$ to use in the
Schr\"{o}dinger equation.

However, the~etalon picture with $H_{phys}$ cannot be applied in
the general case to QG because one cannot resolve the constraints.
It is believed that the Grassmann variables allow writing the
Lagrangian in a form where there are no
constraints~\cite{Savchenko2004,Shest2004,Vereshkov2013,Shest2018,Shestakova2019}.
Using non-canonical gauge fixing~\cite{ruffini2005,Kaku2012} leads
to
\be
Z=\int e^{i\int\left(\pi_{\phi}\phi^\prime -p_a
a^\prime-N\left(-\frac{1}{2}p_a^2+\frac{\pi_{\phi}^2}{2a^2}\right)\right)d\eta}\Pi_\eta
\frac{\delta F}{\delta \varepsilon}\Pi_\eta \delta(F)\mathcal D
p_a \mathcal D a\mathcal D \pi_\phi \mathcal D \phi\mathcal D N,
\label{outin}
\ee
where $F(N)$ is a gauge-fixing~function.

The action (\ref{lag1}) is invariant under the infinitesimal gauge
transformation:
\bea
\tilde a=a+\delta a=a+\varepsilon\, a^\prime,\\
\tilde \phi=\phi+\delta \phi=\phi+\varepsilon\, \phi^\prime,\\
\tilde N=N+\delta N=N+(N\varepsilon)^\prime,\label{NN}
\eea
where $\varepsilon$ is an infinitesimal  function of time. If~one
takes the differential gauge condition $F=N^\prime=0$, then
(\ref{NN}) follows in
\be
\delta F=\delta N^\prime=(N \varepsilon)^{\prime\prime},
\ee
and the Faddev--Popov determinant~\cite{Kaku2012} takes the form
of $\Delta_{FP}=\frac{\delta F}{\delta \varepsilon}=\frac{\delta
N^\prime}{\delta
\varepsilon}=N^{\prime\prime}+2N^\prime\frac{\ptl}{\ptl\eta}+N\frac{\ptl^2}{\ptl\eta^2}$.
The functional (\ref{outin}) could be rewritten as\vspace{-9pt}

\bea
Z=\int e^{i\int\left(\pi_{\phi}\phi^\prime -p_a
a^\prime-N\left(-\frac{1}{2}p_a^2+\frac{\pi_{\phi}^2}{2a^2}\right)-
\bar\theta (N \theta)^{\prime\prime}\right)d\eta}\Pi_\eta
\delta(N^\prime(\eta))\mathcal D N \mathcal D p_a \mathcal D
a\mathcal D \pi_\phi \mathcal D\phi\mathcal D \theta\mathcal D\bar
\theta=\nonumber\\
\int e^{i\int\left(\pi_{\phi}\phi^\prime -p_a
a^\prime-N_0\left(-\frac{1}{2}p_a^2+\frac{\pi_{\phi}^2}{2a^2}-
\bar\theta^\prime  \theta^{\prime}\right)\right)d\eta}
 \mathcal D p_a \mathcal D a\mathcal D
\pi_\phi \mathcal D\phi \mathcal D \theta\mathcal D\bar
\theta=\nonumber\\
\int e^{i\int\left(\pi_{\phi}\phi^\prime -p_a a^\prime+\bar
\theta^\prime\pi_{\bar
\theta}+\pi_\theta\theta^\prime-N_0\left(-\frac{1}{2}p_a^2+\frac{\pi_{\phi}^2}{2a^2}+
\pi_\theta  \pi_{\bar\theta}\right)\right)d\eta}
 \mathcal D p_a \mathcal D a\mathcal D
\pi_\phi \mathcal D\phi \mathcal D \pi_\theta\mathcal D\pi_{\bar
\theta}\mathcal D \theta\mathcal D\bar \theta , ~~~~\label{actg1}
\eea

where using the Grassmann variables~\cite{Kaku2012} in the first
equality of (\ref{actg1}) raises the Faddeev--Popov determinant
into an exponent. Here, a Grassmann number $\bar \theta$ is
considered as a complex conjugate to $\theta$. Integration over
$N$ has been performed explicitly in (\ref{actg1}). If~$N(\eta)$
is discretized  over the interval $\Delta \eta$, the~term
containing a product of the delta functions
$\Pi_\eta\delta(N^\prime(\eta))$ takes the form

\be
\int\dots \delta\left(\frac{N_0-N_1}{\Delta
\eta}\right)\delta\left(\frac{N_1-N_2}{\Delta \eta}\right)\ldots
\delta\left(\frac{N_{k-1}-N_{k}}{\Delta \eta}\right)dN_1 \dots
dN_{k-1}\sim \Delta \eta^{k-1}\delta(N_0-N_k),
\ee

i.e., an~initial value of $N_0$ has to equal a final value $N_k$,
 for instance, one may take  $N_0=1$. For~deducing third equality of  (\ref{actg1}), see Appendix \ref{thetaint}.    Extracting
the Lagrangian from  (\ref{actg1}) gives
\be
L=\pi_{\phi}\phi^\prime -p_a a^\prime+\bar \theta^\prime\pi_{\bar
\theta}+\pi_\theta\theta^\prime-\left(-\frac{1}{2}p_a^2+\frac{\pi_{\phi}^2}{2a^2}+\pi_
\theta\pi_{\bar \theta}\right).
\label{lagun1}
\ee

The action (\ref{lagun1}) is a fixed gauge action with no
Hamiltonian constraint, but~instead, the~ghost (Grassmann)
variables arise in (\ref{lagun1}).

Following Vereshkov and
Shestakova~et~al.~\cite{Savchenko2004,Vereshkov2013}, one may
consider the Hamiltonian
\be
H=-\frac{1}{2}p_a^2+\frac{\pi_{\phi}^2}{2a^2}+\pi_\theta\pi_{\bar
\theta}
\label{ham2}
\ee
as describing the quantum evolution of a~system.

To quantize the system, the~anticommutation relation has to be
introduced for the Grassmann~variables

\be
\{\pi_\theta,\theta\}=-i,~~~~\{\pi_{\bar \theta},\bar \theta\}=-i.
\ee
 {In the particular }representation $\alpha=\ln a$, $\hat
p_\alpha=i\frac{\ptl}{\ptl \alpha}$, $\hat \phi=i\frac{\ptl}{\ptl
k}$, $\hat \pi_\phi=k$, $\hat \pi_\theta=-i\frac{\ptl}{\ptl
\theta}$, $\hat \pi_{\bar \theta}=-i\frac{\ptl}{\ptl \bar
\theta}$, the~Schr\"{o}dinger equation reads as
\be
i\frac{\ptl}{\ptl
\eta}\psi=\left(\frac{1}{2}e^{-2\alpha}\left(\frac{\ptl^2}{\ptl
\alpha^2}+k^2 \right)-\frac{\ptl}{\ptl \theta}\frac{\ptl}{\ptl
\bar \theta}\right)\psi,
\label{shr}
\ee
where the operator ordering in the form of the two-dimensional
Laplacian~\cite{Dew}   has been used\footnote{For further
discussion of the operator ordering issue, see,
e.g.,~\cite{tagirov2001ordering, Pavsic2003}.}. It should be
supplemented by the scalar product
\be
<\psi_1|\psi_2>=\int \psi_1^*(\eta,k,\alpha,\bar \theta ,\theta)
\psi_2(\eta,k,\alpha,\bar \theta ,\theta)e^{2\alpha} d \alpha dk d
\theta d \bar \theta,
\label{35}
\ee
where the measure $e^{2\alpha}$ arises due to the hermicity
requirement~\cite{Dew,Faddeev}. This measure is a consequence of a
minisuperspace metric if the classical  Hamiltonian is written in
the form of $H=\frac{1}{2} {\mathcal
G}^{ij}p_i\,p_j+\pi_\theta\pi_{\bar \theta}$ with $
p_i\equiv\{p_\alpha,\pi_\phi\}$, $\mathcal
G^{ij}=\mbox{diag}\{-e^{-2\alpha},e^{-2\alpha}\}$. Thus, the~
measure takes the form $\sqrt{\mathcal G}=e^{2\alpha}$, $\mathcal
G=|\det\mathcal G_{ij}|$ and the Laplacian
$\frac{1}{\sqrt{\mathcal G}}\frac{\ptl}{\ptl q^i}\sqrt{\mathcal
G}{\mathcal G}^{ij}\frac{\ptl}{\ptl
q^i}=e^{-2\alpha}\left(-\frac{\ptl^2}{\ptl \alpha^2}+\pi_\phi^2
\right)$ is self-adjoined~\cite{Dew} with this measure. Formal
solutions of the Equation~(\ref{shr}) can be written as
\be
\psi(\eta,k,\alpha,\bar\theta,\theta)=(\bar\theta+\theta)u(\eta,k,\alpha)+i(\bar
\theta-\theta)v(\eta,k,\alpha),
\ee
where the functions  $u$ and $v$ satisfy the equation
\be
i\frac{\ptl}{\ptl \eta}u=\hat H_0 u
\label{shr0}
\ee
with
\be
\hat H_0=\frac{1}{2}e^{-2\alpha}\left(\frac{\ptl^2}{\ptl
\alpha^2}+k^2 \right).
\label{H0s}
\ee

 Then, the~scalar product (\ref{35}) reduces to
\be
<\psi_1|\psi_2>=-2i\int \left(u_1^*v_2-v_1^*u_2\right)e^{2\alpha}
d \alpha dk.
\label{norm}
\ee

Although the constraint $H_0=0$ formally disappears from the
theory, one may think that the space of solutions of the
Wheeler--DeWitt equation (WDW) still plays a role~\cite{Mont}.
Otherwise, the~question of correspondence with the classical
theory, where the Hamiltonian constraint holds, arises. We would
like to relate the space of the functions, satisfying the
Schr\"{o}dinger Equation~(\ref{shr}) with the functions $\chi$
satisfying the equation $H_0\chi=0$, i.e.,~the WDW equation. The~
operator $\hat H_0$ (\ref{H0s}) has the Klein--Gordon form. Thus,
the Klien--Gordon-type scalar product has to be used. According to
this hypothesis, let us represent the functions $u$, $v$ as
\bea
v(\alpha,k)=e^{-iH_0\eta}\hat D^{1/4}\chi(\alpha,k),\label{vv}\\
u(\alpha,k,\eta)=e^{-iH_0\eta}\hat
D^{-1/4}\delta(\alpha-\alpha_0)\frac{\ptl}{\ptl
\alpha}\chi(\alpha,k),\label{uu}
\eea
where the operator  $\hat D=-\frac{\ptl^2}{\ptl\phi^2}$, or~
$D=k^2$ in the representation (\ref{repr}) and
\be
\chi(\alpha,k)=\frac{e^{-i\,\alpha|k|-\alpha_0}}{\sqrt{2|k|}}C(k),
\label{neg1}
\ee
and, as~in  (\ref{psi}),  only a half-space corresponding to the
negative frequencies' solutions of the WDW equation is taken
because only in this case does the Klein--Gordon product imply a
positive definite norm of a state. The~operator $\hat D$ (see
Appendix in~\cite{mostf}) is a necessary attribute of the scalar
product for the Klein--Gordon equation to obtain hermicity. It
should be noted that in~fact, the~function $v$ does not depend on
the time $\eta$ because $\hat H_0\chi=0$ and $\hat D$ commutes
with $H_0$. Thus, the time evolution arises only due to function
$u$, or~more accurately, due to the presence of the Dirac delta
function in (\ref{uu}).

 Thus, the scalar product (\ref{norm}) reduces
to
\be
<\psi_1|\psi_2>=-2i\int \left(\frac{\ptl \chi_1^*}{\ptl
\alpha}\chi_2-\chi_1^*\frac{\ptl \chi_2}{\ptl
\alpha}\right)e^{2\alpha}\biggr|_{\,\alpha=\alpha_0} dk.
\label{norm1}
\ee

\noindent  {The expression for} the mean value of an operator
$\hat A$ has the form:
\be
  <\psi|\hat A|\psi>=-2i\int e^{2\alpha}\biggl( u^*\hat A v-v^*\hat
  A u \biggr)\biggr|_{\,\alpha=\alpha_0\rightarrow-\infty}
  dk,\label{mean3}
\ee
where $u$, $v$ are given by (\ref{vv}) and (\ref{uu}), and it is
assumed that an  operator $\hat A$ does not contain the ghost
variables $\theta$, $\bar \theta$, that is expected for physical
operators. The~limit $\alpha\rightarrow -\infty$ in (\ref{mean3})
implies that an evolution begins at $\eta=0$ when $a=0$ and
$\alpha=\ln a$ tends to $-\infty$.

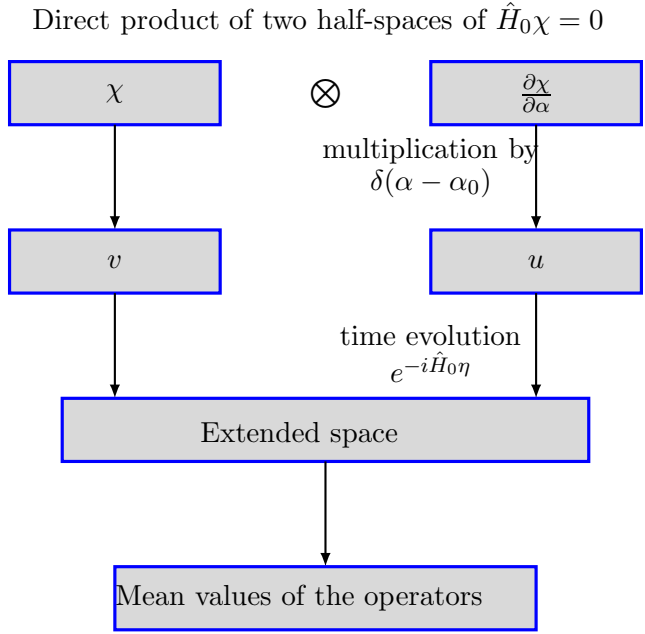
\begin{figure}[H]

\begin{tikzpicture}
\draw[blue, very thick,fill=gray!30] (0+0.7*2,0+0.7*10) rectangle
(0.7*4+0.7*2,0.7*1.2+0.7*10); \draw[blue, very thick,fill=gray!30]
(0+0.7*2+0.7*8,0+0.7*10) rectangle
(0.7*4+0.7*2+0.7*8,0.7*1.2+0.7*10);
 \draw (0+0.7*2+0.7*2,0.7*10+0.7*0.6) node {$\chi$};
  \draw (0+0.7*2+0.7*2+0.7*8,0.7*10+0.7*0.6) node {$\frac{\ptl\chi}{\ptl\alpha}$};
  \draw (0+0.7*2+0.7*2+0.7*4,0.7*10+0.7*0.6) node {$\bigotimes$};
  \draw (5.5,0.7*10+0.7*0.6+1) node {Direct product of two half-spaces of $\hat H_0\chi=0$};
    \draw[->,  thick,   arrows={-latex}]  (0.7*4,0.7*10) -- (0.7*4,0.7*8);
    \draw[->,  thick,   arrows={-latex}]  (0.7*4+0.7*8,0.7*10) -- (0.7*4+0.7*8,0.7*8);
    \draw (0+0.7*10,0.7*8.9+0.7*0.6) node {multiplication by};
   \draw (0+0.7*10,0.7*8.9) node {$\delta(\alpha-\alpha_0)$};
   \draw[blue, very thick,fill=gray!30] (0+0.7*2,0+0.7*10-0.7*3.2) rectangle
(0.7*4+0.7*2,0.7*1.2+0.7*10-0.7*3.2); \draw
(0+0.7*2+0.7*2,0.7*10+0.7*0.6-0.7*3.2) node {$v$}; \draw[blue,
very thick,fill=gray!30] (0+0.7*2+0.7*8,0+0.7*10-0.7*3.2)
rectangle (0.7*4+0.7*2+0.7*8,0.7*1.2+0.7*10-0.7*3.2); \draw
(0+0.7*2+0.7*2+0.7*8,0.7*10+0.7*0.6-0.7*3.2) node {$u$};
 \draw[->,  thick,   arrows={-latex}]  (0.7*4,0.7*10-0.7*3.2) -- (0.7*4,0.7*8-0.7*3.2);
    \draw[->,  thick,   arrows={-latex}]  (0.7*4+0.7*8,0.7*10-0.7*3.2) -- (0.7*4+0.7*8,0.7*8-0.7*3.2);
     \draw (0+0.7*10,0.7*8.9+0.7*0.6-0.7*3.5) node {time evolution};
   \draw (0+0.7*10,0.7*8.9-0.7*3.5) node {$e^{-i\hat H_0\eta}$};
   \draw[blue, very thick,fill=gray!30] (0+0.7*3,0+0.7*3.6) rectangle
(0.7*8+0.7*5,0.7*1.2+0.7*3.6);
 \draw (0.7*7.5,0.7*4.1) node {Extended space};
  \draw[->,  thick,   arrows={-latex}]  (0.7*8,0.7*3.6) -- (0.7*8,0.7*1.6);
   \draw[blue, very thick,fill=gray!30] (0+0.7*4,0+0.7*0.4) rectangle
(0.7*8+0.7*4,0.7*1.2+0.7*0.4);  \draw (0.7*7.5,0.7*1) node {Mean
values of the operators};
\end{tikzpicture}

\caption {Schematic illustration of the scalar product
(\ref{vv})--(\ref{mean3}).}
\label{fig31}
\end{figure}

The evaluation of the mean values is illustrated schematically in
Figure \ \ref{fig31}. Initially, we start with the negative
frequency functions $\chi$, $\frac{\ptl\chi}{\ptl \alpha}$ both
satisfying $H_0\chi=0$. A~direct product of these half-spaces is
taken. Then, the function $\frac{\ptl\chi}{\ptl \alpha}$ is
multiplied by $\delta(\alpha-\alpha_0)$ and runs into an extended
space, where ``evolution'' occurs, and, thus, the~mean values of
the operators can be~evaluated.

Both Schr\"{o}dinger and Heisenberg pictures are possible with
this scalar product. In the Heisenberg picture, the~time-dependent
operators have the form
\be
\hat A(\eta)=e^{i\hat H_0\eta}\hat Ae^{-i\hat H_0\eta},
\ee
while the functions $u$ and $v$ have to be used without multiplier
$e^{-i\hat H_0\eta}$.

\section{Expectation Values of Scale Factor~Degrees}

The simplest way to test a theory is to compare it with the etalon
picture by calculating the mean value of the squared scale factor,
which has to be equal to $\frac{16}{3\sqrt{\pi}}\eta$ according to
(\ref{resa2}).  To~do that, it is sufficient to expand $e^{-i\hat
H_0\eta}\approx 1-i\hat H_0\eta -\frac{1}{2}\hat H_0^2\eta^2$ in
(\ref{vv}) and (\ref{uu}) and perform the calculation according
(\ref{mean3}). It turns out that the mean value of
$a^2=e^{2\alpha}$ actually coincides with that given by
(\ref{resa2}). The~next test is the calculation of $<C|a^4|C>$.
The result of the calculation is
\be
<C|a^4|C>=2\eta^2,
\label{a4}
\ee
while the etalon model gives another value (\ref{resa4}). The
origin of this discrepancy could be better seen in the Heisenberg
picture. Evolution equations for the Heisenberg operators follow
from the operator commutators with the Hamiltonian (\ref{H0s})
\be
\frac{d \hat a^2}{d\eta}=i[\hat H_0,\hat a^2].
\label{eqcomm}
\ee
 {It is possible to guess} a solution for this particular case:
\be
\hat a^2(\eta)=e^{2\alpha}+2\eta \,e^{-\alpha}\hat p_\alpha
e^{\alpha}-2\eta^2\hat H_0,
\label{sola2}
\ee
where we define $\hat p_\alpha=i\frac{\partial}{\partial
\alpha}$\footnote{Instead, one could define self-ajoind $\hat
p_\alpha=i{\mathcal G}^{-1/4}\frac{\ptl}{\ptl\alpha}{\mathcal
G}^{1/4}=i\left(\frac{\ptl}{\ptl\alpha}+1\right)$ \cite{Dew} and
rewrite the Equations~(\ref{eqcomm}) and (\ref{sola2}) using this
definition.}.

\begin{figure}[H]
  \includegraphics[width=9cm]{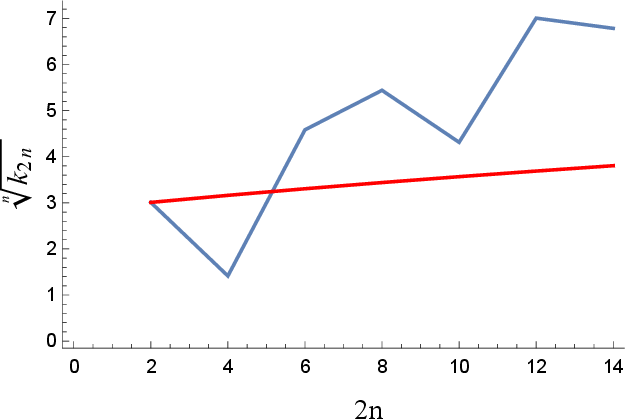}\\
  \caption{ {n}
-th root of coefficient $k_{2n}$ in the expression $<C|\hat
a^{2n}|C>=k_{2n}\,\eta^n$
  for the mean value of the $2n$th degree of scale factor with respect to the wave packet (\ref{pak}).
  The red and blue curves correspond to the etalon  method and that with the Grassmann variables,
  respectively.
   }\label{fig1}
\end{figure}

Actually, the~calculation of the commutator  (\ref{eqcomm}) using
(\ref{H0s}), (\ref{sola2}) gives
\be
i[\hat H_0,\hat a^2(\eta)]=2 \,e^{-\alpha}\hat p_\alpha
e^{\alpha}-4\eta \hat H_0,
\ee
which is exactly equal to the derivative of (\ref{sola2}) over
$\eta$. Under~calculation of the mean value of $<C|\hat a^2|C>$,
the third term in (\ref{sola2}) does not contribute, and~the
result coincides with that of the etalon method. However,
under~the calculation of $<C|\hat a^4|C>$, the~first and third
terms in (\ref{sola2}) play a role, and the discrepancy with the
etalon method arises. One can calculate the mean values of the
other degrees of $a$, which are presented in Table~\ref{tab}. It
is interesting to plot the values of
$\sqrt[n]{k_{2n}}=\frac{1}{\eta}\sqrt[n]{<C|\hat
a^{2n}(\eta)|C>}$, which is shown in Figure~\ref{fig1}.

\begin{table}[H]
\caption{\label{nn1}%
The expectation values $<C|a^{2n}|C>=k_{2n}\,\eta^n$ for the wave
packet (\ref{pak}). }
 \begin{tabular}{cccccccc}
\toprule
\boldmath{$2n$}   & \textbf{2}         & \textbf{4}& \textbf{6}&\textbf{8}&\textbf{10}&\textbf{12}&\textbf{14} \\
\hline $\sqrt[n]{k_{2n}}$ for the etalon model &$\frac{16}{3
\sqrt{\pi}}$         & $\sqrt{10}$& $\sqrt[3]{\frac{64}{\sqrt\pi}
}$& $ \sqrt[4]{140}$ & $ \sqrt[5]{\frac{1024}{\sqrt\pi}}$
&$\sqrt[6]{2520}$&$ \sqrt[7]{\frac{20480}{\sqrt\pi}}$\\
$\sqrt[n]{k_{2n}}$ for the model with the Grassmann variables &
$\frac{16}{3 \sqrt{\pi }}$ &$\sqrt{2}$&
$\sqrt[3]{\frac{512}{3\sqrt\pi}} $&$\sqrt[4]{876}$& $
\sqrt[5]{\frac{7936}{3\sqrt\pi}}$& $
\sqrt[6]{118280}$& $ \sqrt[7]{\frac{1172480}{\sqrt\pi}}$\\
\hline
 \end{tabular}
\label{tab}
\end{table}
\unskip

Recently, analog models of QG and~minisuperspace have been
 discussed~\cite{universe8090455,universe8090482}. In~
Appendix \ref{particle}, quantization of a particle-clock is
considered, which has some similar features to the minisuperspace
model but~does not have an operator ordering~issue.

\section{Discussion and~Conclusions}

A reasonable expression for the scalar product using the Grassmann
variables is suggested. It establishes a relation of a picture
with the Grassmann variables to the Klein--Gordon scalar product
and allows calculating the mean values of operators in both
Schr\"{o}dinger and Heisenberg pictures, which give the same
results. However, it is shown that the mean values of $a^{2n}$ are
different for $n>1$ than those calculated in the etalon method,
implying an explicit exclusion of the superfluous degrees of
freedom. One may guess that the above methods could have different
Hilbert spaces. That means that the different wave packets
 have to be taken for these methods to obtain the same set of operator mean values. Here, we cannot find a wave
packet $\tilde C(k)$, which would give the same mean values as a
wave packet $C(k)$ for the etalon~method.

 The possible influence of   the Zitterbewegung phenomenon in
 extended space was investigated in Appendix \ref{zitt} but~without a breakthrough in the results achieved. It should be noted that the quasi-Heisenberg picture corresponds
entirely with the etalon method~\cite{cherkas2020evidence}.

 One of the possible ways to correct the picture with the Grassmann variables is to assume
 that the operators of physical observables act not only in $k$ and
 $\alpha$ space but~also in the extended space of the Grassmann variables $\theta$,
 $\bar \theta$.  This hypothesis needs further investigation\footnote{In this relation, see~\cite{Chauhan2022,shukla2023antibrst}, where an auxiliary pair of the Grassmann variables is introduced.} as well
 as the general issue of the scalar product
for the approach with the Grassmann~variables.

\appendix
\section{Generalized Hamiltonian
 Form of the Action with the Grassmann~Variables}
\label{thetaint}
Let us  prove the equivalence of the action functional with the
Grassmann variables in the generalized Hamiltonian form to the
conventional action in the Lagrangian form. Stating from
$Z_{\theta}$ in the generalized Hamiltonian form, let us use the
invariance of the continual integral relative change of the
variables~\cite{Kaku2012}. Implementing the  change
$\pi_\theta=\xi+\bar \theta^\prime$ and $\pi_{\bar \theta}=\bar
\xi+\theta^\prime$, we~have
\bea
Z_\theta=\int e^{i\int \left(\bar \theta^\prime\pi_{\bar
\theta}+\pi_\theta\theta^\prime-\pi_{
\theta}\pi_{\bar\theta} \right)d\eta}\mathcal D \pi_\theta\mathcal D\pi_{\bar\theta}\mathcal D \theta\mathcal D\bar \theta=\nonumber\\
\int e^{i\int \left(\bar
\theta^\prime(\bar\xi+\theta^\prime)+(\xi+\bar\theta^\prime)\theta^\prime
-(\xi+\bar\theta^\prime)(\bar\xi+\theta^\prime)\right)d\eta}\mathcal D \xi\mathcal D\bar \xi\mathcal D \theta\mathcal D\bar \theta=\nonumber\\
\int e^{i\int \bar \xi\xi d\eta}\mathcal D \xi\mathcal D\bar
\xi\int e^{i\bar\theta^\prime  \theta^{\prime} d\eta}\mathcal D
\theta\mathcal D\bar \theta\sim \int e^{i\bar\theta^\prime
\theta^{\prime} d\eta}\mathcal D \theta\mathcal D\bar \theta,
\eea
where a multiplier containing the integral over $\xi$, $\bar \xi$
is omitted in the last~equality.

Certainly, another way to prove the equivalence is to find the
momentums by varying the action
\be
S_\theta=\int \left(\bar \theta^\prime\pi_{\bar
\theta}+\pi_\theta\theta^\prime-\pi_ \theta\pi_{\bar
\theta}\right)d\eta
\label{stheta}
\ee
over $\pi_\theta$, $\pi_{\bar \theta}$. That gives
\be
\pi_\theta=\bar \theta^\prime,~~~\pi_{\bar
\theta}=\theta^\prime.~~~~
\label{momentums}
\ee
Then, substitution of the momentums (\ref{momentums}) into
(\ref{stheta}) results in
\be
S_\theta=\int \bar\theta^\prime  \theta^{\prime}d\eta.
\ee

\section{Quantization of a~Particle-Clock}
\label{particle}

\begin{figure}[H]
\includegraphics[width=10cm]{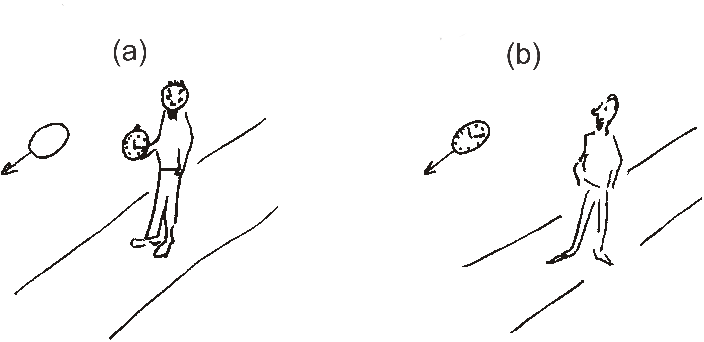}
\caption{(\textbf{a}) Observer describing a particle motion by his
clock, (\textbf{b}) description of a particle motion by ``particle
clock''.}
\label{stoks}
\end{figure}

It is interesting to consider the  scalar product introduced above
by giving an example of a relativistic particle, having its own
clock (see Figure~\ref{stoks}). It could be a radioactive particle
decaying exponentially with a probability
\be
P(\tau)=\frac{1}{T}e^{-\tau/T},
\label{prob}
\ee
where $T$ is a mean lifetime of the particle. We will consider
$\tau$ as a proper time of a~system. \vspace{-12pt}

The action of a relativistic particle  can be defined
as~\cite{Kaku2012}:
\begin{equation}
S=\frac{1}{2}\int(e^{-1}\dot{x}_\mu^2-e \,m^2)d\tau,  \label{s2}
\end{equation}
 where the $x^\mu=\{t,\bm x\}$, $(-,+,+,+)$
 signature is used, and the lapse function $e(\tau)$ is
 introduced.
 One more equivalent form resulting in
 (\ref{s2}) after varying over $p_\mu$ looks as
\begin{equation}
S=\int\left(p_\mu
\dot{x}^\mu-e\left(\frac{p_\mu^2+m^2}{2}\right)\right)
d\tau=\int\left(\bm p \bm x^\prime-p_t t^\prime
-\frac{e}{2}\left(-p_t^2+\bm p^2+m^2\right)\right) d\tau.
\label{partaction}
\end{equation}

From (\ref{partaction}), it follows that the particle analog of
the
 minisuperspace Hamiltonian (\ref{phihamil}) is written as
\be
H=\frac{e}{2}(-p_t^2+\bm p^2+m^2),
\ee
and it is constraint $\Phi_1$ simultaneously. The equations of
motion are
\bea
\frac{d\bm x}{d \tau}=\frac{\ptl H}{\ptl \bm p}={e}\bm
p,~~~~~~~\bm p=const,\nonumber\\
\frac{d t}{d \tau}=-\frac{\ptl H}{\ptl p_t}={e} p_t, ~~~~~~~~~~
p_t=const.
\eea

The additional, depending on $\tau$, gauge-fixing condition
\be
\Phi_2=t-\frac{\sqrt{\bm p^2+m^2}}{m}\,\tau=0,
\ee
  assigns $e=1/m$ and after calculations (\ref{hpwf}), (\ref{testeqmov}) leads
to the physical Hamiltonian
\be
H_{phys}=p_t \,t^\prime+\frac{d F(\bm p,\tau)}{d\tau}=\frac{\bm
p^2+m^2}{m}+\frac{\ptl F(\bm p,\tau)}{\ptl \tau}
\label{neweq}
\ee
describing a particle motion in the proper time $\tau$. In~
Equation~(\ref{neweq}), the total derivative is changed by the
partial derivative because momentum $\bm p$ does not depend on
time. On the other hand, the physical Hamiltonian must reproduce
motion in the reduced space to give
\be
\bm x^\prime = \frac{\ptl H_{phys}}{\ptl \bm p}=\frac{\bm p}{m},
\ee
thus, $\frac{\ptl F(\bm p,\tau)}{\ptl \tau}=-\frac{\bm
p^2+m^2}{2m}$ and
\be
H_{phys}=\frac{\bm p^2+m^2}{2m}.
\ee
Analogously to (\ref{sh})--(\ref{hata}), a quantum picture with
the Grassmann variables leads to the Schr\"{o}dinger equation
\be
i\frac{\ptl}{\ptl
\eta}\psi=\left(\frac{1}{2m}\left(\frac{\ptl^2}{\ptl t^2}+\bm
p^2+m^2 \right)-\frac{\ptl}{\ptl \theta}\frac{\ptl}{\ptl \bar
\theta}\right)\psi.
\label{shr3}
\ee
The mean value of the operator $\hat A(t,i\frac{\ptl}{\ptl t},\bm
p,i\frac{\ptl}{\ptl \bm p})$ has the form:
\be
  <\psi|\hat A|\psi>=-2i\int \biggl( u^*\hat A v-v^*\hat
  A u \biggr)\biggr|_{t=t_0\rightarrow 0}
  d^3\bm p,\label{mean3t}
\ee
where $u$, $v$ are given by
\bea
v(t,\bm p)=D^{1/4}\chi(t,\bm p),\label{vv11}\\
u(t,\bm p,\tau)=e^{-iH_0 \tau}
D^{-1/4}\delta(t-t_0)\frac{\ptl}{\ptl t}\chi(t,\bm p),\label{uu11}
\eea
\be
H_0=\frac{1}{2m}\left(\frac{\ptl^2}{\ptl t^2}+\bm p^2+m^2 \right),
\ee
$D=\varepsilon^2(p)= {m^2+\bm p^2}$, and~
\be
\chi(t,\bm p)=\frac{e^{-i\,\varepsilon
t}}{\sqrt{2\varepsilon}}C(\bm p).
\label{neg}
\ee

For the wave packet
\be
C(\bm p)\sim e^{-(\bm p-\bm p_0)^2/(2\sigma^2)},
\ee
the mean value of $<\psi|t^2|\psi>$ equals
\be
<\psi|t^2|\psi>=\frac{\tau ^2 \left(\varepsilon_0^2+3 \sigma
^2\right)}{m^2},
\ee
where $\varepsilon_0^2=p_0^2+m^2$. It turns out to be the same for
both methods: the physical Hamiltonian and that with the Grassmann
variables. Calculation of the mean value of $<\psi|t^4|\psi>$
 gives
\be
<\psi|t^4|\psi>=\frac{\tau ^4 \left(\varepsilon_0^4+2 \sigma^2
\left(5 \varepsilon_0^2-2m^2\right)+15 \sigma ^4\right)}{m^4}
\label{t41}
\ee
for the physical Hamiltonian method and~
\be
<\psi|t^4|\psi>=\frac{\tau ^4 \left(\varepsilon_0^4+2 \sigma ^2
\left(5 \varepsilon_0^2-2m^2\right)+15 \sigma
^4\right)}{m^4}-\frac{3 \tau ^2}{m^2}
\label{t42}
\ee
for the method with the Grassmann variables. Compared to
Equation~(\ref{t41}), the~additional term $\frac{3 \tau ^2}{m^2}$
appears in (\ref{t42}). The~value of (\ref{t42})  averaged over
probability  of particle decay (\ref{prob}) leads to a quantity,
which could be, in~principle, observed experimentally:

\vspace{-6pt}
\be
\int_0^\infty<\psi|t^4|\psi>P(\tau)d\tau= \frac{24T^4
\left(\varepsilon_0^4+2 \sigma ^2 \left(5
\varepsilon_0^2-2m^2\right)+15 \sigma ^4\right)}{m^4}-\frac{6
T^2}{m^2}.
\label{expobs}
\ee
The last term in (\ref{expobs}) becomes considerable when the
particle width $\Gamma=1/T$ is comparable with the particle~mass.

\section{Removing of an ``Extended~Zitterbewegung''}
\label{zitt}

The well-known phenomenon of  Zitterbewegung
(see~\cite{Silenko2020,Lovett2023} and references therein) is an
inevitable feature of any relativistic field equation and is
usually removed by the Foldy--Wouthuysen
transformation~\cite{Costella1995,Neznamov2009,Silenko2016}. It
arises due to interference of the solutions of the Klein--Gordon
equation with the positive and negative frequencies. Here, we
discuss the solution of the Schr\"{o}dinger Equation~(\ref{shr0})
with the WDW operator on the right-hand side and will  consider a
possible ``extended Zitterbewegung''. In the extended space,
the~solutions of the Schr\"{o}dinger Equation~(\ref{shr0}) look as
$u_\lambda=e^{-i\lambda \eta}\psi_\lambda$ where the
eigenfunctions $\psi_\lambda$ satisfy
\be
H_0\psi=\lambda\, \psi_\lambda
\label{eagen}
\ee
with a different sign of $\lambda$. For~$\lambda>0$, the general
solution takes the form
\be
\psi_\lambda=c_1 J_{i |k|}(\sqrt{2\lambda} \,e^{\alpha})+c_2 J_{-i
|k|}(\sqrt{2\lambda}\, e^{\alpha}),
\ee
where $J_{i |k|}$ is the Bessel function of an imaginary index.
The~ solutions for $\lambda<0$ were investigated
in~\cite{Gryb2019,Gielen2020,Menendez,Gielen2022}.

 The solution $J_{-i |k|}(\sqrt{2\lambda}
e^{\alpha})$ in the extended space is ``an heir'' of the mass
shell solution for a negative frequency (\ref{neg}) by virtue of
\be
\lim_{\lambda\rightarrow 0}\left(\frac{\lambda}{2}\right)^{i
|k|/2}J_{-i |k|}(\sqrt{2\lambda} e^{\alpha})= e^{-i
|k|\alpha}/\Gamma(1-i |k|),
\ee
where $\Gamma$ is a Gamma function. The~function
$\delta(\alpha-\alpha_0)\frac{\ptl}{\ptl \alpha}\chi(\alpha,k)\sim
\delta(\alpha-\alpha_0) e^{-i|k|\alpha} $ in (\ref{uu}) is a
superposition of the extended space functions for both $\lambda>0$
and $\lambda<0$.

Let us take an alternative expression
\be
u(\alpha,k,\eta)=e^{-i\hat H_0\eta} \frac{ e^{-i \alpha  |k|} \,
_1{F}_1\left(\frac{3}{2};1-i |k|;-\frac{e^{2 \alpha }}{2 \gamma
}\right)}{2 \sqrt{2} \gamma \sqrt{|k|} }C(k),
\label{inste}
\ee
which consists of only the superposition of  eigenfunctions with
$\lambda
>0$:
\be
\int_0^\infty
e^{-\gamma\lambda}\sqrt{\lambda}(2\lambda)^{i|k|/2}J_{-i|k|}(\sqrt{2\lambda}\,e^{2\alpha})d\lambda=\frac{\sqrt{\pi
}\, 2^{i |k|} e^{-i|k|\alpha} \, _1{F}_1\left(\frac{3}{2};1-i
|k|;-\frac{e^{2\alpha}}{2 \gamma }\right)}{2\gamma
^{3/2}\Gamma(1-i|k|)},
\label{uu1}
\ee
where $_1{F}_1$ is a confluent hypergeometric function, and
$\gamma$ is some positive parameter. It should be noted that the
superposition (\ref{uu1}) contains the functions of the extended
space $J_{-i|k|}$ corresponding to the $e^{-i|k|\alpha}$ functions
(\ref{neg}) on an on-shell space but~not the functions $J_{i|k|}$
referring to the positive frequency solutions $e^{i|k|\alpha}$ of
the WDW~equation.

When $\gamma$ tends to zero, the~function
$\frac{_1{F}_1\left(\frac{3}{2};1-i |k|;-\frac{a^2}{2 \gamma
}\right)}{2\gamma }$ peaks near $a=0$, i.e.,~near $\alpha=\ln a\to
-\infty$. In addition,~$\frac{\Gamma (-i |k|)}{2 \gamma \Gamma
(1-i |k|) }\int_{0}^\infty
 \, _1{F}_1\left(\frac{3}{2};1-i
|k|;-\frac{a^2}{2 \gamma }\right)ada=1$; thus, the~limit
$\gamma\rightarrow 0$ is an analog of using
$\delta(\alpha-\alpha_0)$ in (\ref{uu}) and tending
$\alpha_0\rightarrow -\infty$.

Calculations of the mean value of $a^2$  using (\ref{inste}) gives
\begin{align}
<C|\hat a^2|C>=4 \gamma+\Aboxed{\frac{16 \eta }{3 \sqrt{\pi
}}}+\frac{\eta ^2}{2\gamma}.
\end{align}
As one can see, a more complicated regularization is needed,
because the limit $\gamma\rightarrow 0$ gives infinity and we need
to extract the terms which do not depend on $\gamma$. The~
situation is similar to that in~\cite{cherkas2020evidence} for
this method. After~such a regularization, we have the same mean
value as in (\ref{resa2}). The calculation of $a^4$ gives
\begin{align}
 <C|\hat a^4|C>=-\frac{16
\gamma ^2}{3}+\frac{128 \,\gamma\,  \eta }{\sqrt{\pi }}+\Aboxed{2
\eta ^2}+\frac{16 \eta ^3}{3 \sqrt{\pi } \gamma }+\frac{3 \eta
^4}{2 \gamma ^2}
\end{align}
which after regularization, i.e.,~omitting  the terms depending on
$\gamma$, coincides with (\ref{a4}) but~not with the etalon result
(\ref{resa2}). Thus, removing the possible ``extended
Zitterbewegung'' does not lead to the coincidence with the etalon
picture.

\bibliography{time1}
\end{document}